\documentclass[a4paper,11pt]{article}

\usepackage{jinstpub} 

\usepackage{graphicx}
\usepackage{subcaption}
\usepackage{lineno}
\usepackage{verbatim}
\usepackage{units}

\title{Reconstruction and Measurement of $\mathcal{O}$(100) MeV Energy Electromagnetic Activity from $\pi^0 \rightarrow \gamma\gamma$ Decays in the MicroBooNE LArTPC}

\collaboration{MicroBooNE Collaboration}

\author[j]{C.~Adams}
\author[l]{M.~Alrashed}
\author[k]{R.~An}
\author[c]{J.~Anthony}
\author[cc]{J.~Asaadi}
\author[p]{A.~Ashkenazi}
\author[hh]{S.~Balasubramanian}
\author[i]{B.~Baller}
\author[q]{C.~Barnes}
\author[t]{G.~Barr}
% ...

% ...
\affiliation[c]{University of Cambridge, Cambridge CB3 0HE, United Kingdom}
% ...
\affiliation[i]{Fermi National Accelerator Laboratory (FNAL), Batavia, IL 60510, USA}
\affiliation[j]{Harvard University, Cambridge, MA 02138, USA}
\affiliation[k]{Illinois Institute of Technology (IIT), Chicago, IL 60616, USA}
\affiliation[l]{Kansas State University (KSU), Manhattan, KS, 66506, USA}
% ...
\affiliation[p]{Massachusetts Institute of Technology (MIT), Cambridge, MA, 02139, USA}
\affiliation[q]{University of Michigan, Ann Arbor, MI, 48109, USA}
% ...
\affiliation[t]{University of Oxford, Oxford OX1 3RH, United Kingdom}
% ...
\affiliation[cc]{University of Texas, Arlington, TX, 76019, USA}
% ...
\affiliation[hh]{Wright Laboratory, Department of Physics, Yale University, New Haven, CT, 06520, USA}

  \emailAdd{microboone\_info@fnal.gov}

\abstract{We present results on the reconstruction of electromagnetic (EM) activity from photons produced in charged current $\nu_{\mu}$ interactions with final state $\pi^0$s. We employ a fully-automated reconstruction chain capable of identifying EM showers of $\mathcal{O}$(100) MeV energy, relying on a combination of traditional reconstruction techniques together with novel machine-learning approaches. These studies demonstrate good energy resolution, and good agreement between data and simulation, relying on the reconstructed invariant $\pi^0$ mass and other photon distributions for validation. The reconstruction techniques developed are applied to a selection of $\nu_{\mu} + {\rm Ar} \rightarrow \mu + \pi^0 + X$ candidate events to demonstrate the potential for calorimetric separation of photons from electrons and reconstruction of $\pi^0$ kinematics.}

\begin{document}
\maketitle
\flushbottom

\section{Introduction} 
\par We present studies of electromagnetic (EM) showers from photons produced in the decay of neutral pions ($\pi^0$s) that originate from charged current (CC) $\nu_{\mu}$ interactions recorded with the MicroBooNE detector~\cite{bib:MicroBooNE} on the Booster Neutrino Beamline (BNB)~\cite{bib:BNB} at Fermilab. This work focuses on the reconstruction and characterization of EM showers in the \unit[30-250]{MeV} energy range. Particular emphasis is given to studies of energy reconstruction. This paper describes in detail the employment of a fully-automated reconstruction technique in a liquid argon time projection chamber (LArTPC) for this topology of interactions.

\par Measuring and characterizing the signatures of electromagnetic showers is key to the success of the neutrino oscillation programs of SBN~\cite{bib:SBN} and DUNE~\cite{bib:DUNE}, which both rely on measuring $\nu_e$ appearance in a $\nu_{\mu}$ beam to search for possible sterile neutrinos and perform precision neutrino oscillation measurements, respectively. The topology of electromagnetic activity with energies of a few hundred MeV makes the reconstruction particularly challenging.

\par To date, measurements of EM showers from $\pi^0$ decay photons in a LArTPC have been published by the ICARUS, ArgoNeuT, and MicroBooNE~\cite{bib:ccpi0} collaborations. The ICARUS experiment has made measurements of $\pi^0$s from both cosmic-ray~\cite{bib:ICARUS_pi0_CR} and neutrino~\cite{bib:ICARUS_pi0_NU} interactions, both with samples of order 100 reconstructed $\pi^0$ candidates. The ArgoNeuT collaboration has measured $\pi^0$ decays to perform a measurement of semi-inclusive neutral current (NC) production in the NuMI beamline~\cite{bib:ArgoNeuT_NC}. ArgoNeut's photon energy reconstruction capability is limited by its small volume and lack of containment of electromagnetic showers. The ArgoNeuT collaboration also released the first study on $e$/$\gamma$ separation using calorimetry~\cite{bib:argoneutdedx}. This work expands on previous literature by implementing a fully-automated EM energy reconstruction and presenting detailed studies of energy reconstruction and resolution which describe the various sources of energy smearing and bias. This work uses a sample of $440 \pm 21$ (of which $88$ expected background) candidate $\pi^0$ events, the largest available to date. Accurate and efficient reconstruction of electromagnetic activity in LArTPCs is a key to the success of a broad physics program which aims to perform detailed differential cross section measurments of $\pi^0$ production, precise $\nu_{\mu} \rightarrow \nu_e$ neutrino oscillation measurements, as well as tests of beyond the Standard Model physics models which manifests themselves through $\mathcal{O}$(\unit[100]{MeV}) EM signatures.

\par  Section~\ref{sec:eloss} summarizes energy loss mechanisms for electrons and photons in liquid argon. Section~\ref{sec:shrreco} describes the shower reconstruction employed in this work, and is followed by a description of the event selection applied to obtain $\nu_{\mu}$ CC $\pi^0$ events in section~\ref{sec:pi0selection}. Section~\ref{sec:ereco} is dedicated to shower energy resolution studies, followed by a presentation of results pertaining to measurements of $\pi^0$ and $\gamma$ shower metrics in section~\ref{sec:pi0}. Finally, a brief conclusion is presented in section~\ref{sec:conclusions}.

\section{Electron and Photon Propagation in Argon}
\label{sec:eloss}
\par This section introduces EM energy loss in liquid argon, focusing on the features that lead to the characteristically sparse and stochastic nature of EM showers of $\mathcal{O}(100)$ MeV. This work builds upon previous studies of Michel decay electrons in MicroBooNE~\cite{bib:michel}.

\subsection{Electron Energy Loss}

\par
\begin{figure}[ht]
    \centering
    \begin{subfigure}[b]{0.45\textwidth}
    \centering
    \includegraphics[width=1.00\textwidth]{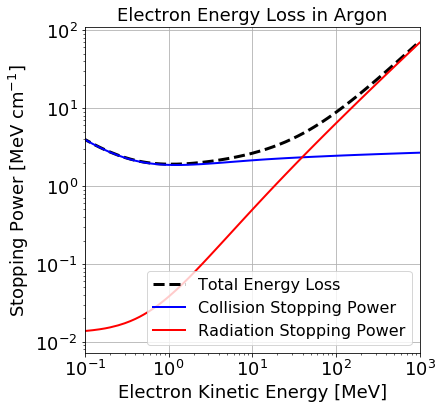}
    \caption{}
    \end{subfigure}
    \begin{subfigure}[b]{0.45\textwidth}
    \centering
    \includegraphics[width=1.00\textwidth]{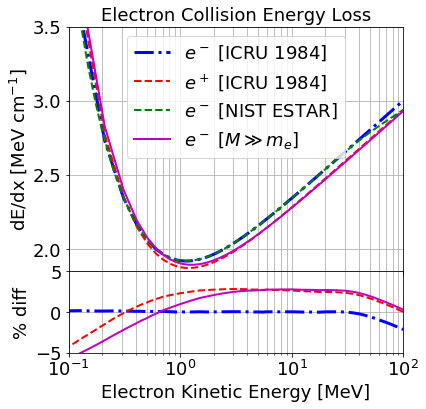}
    \caption{}
    \end{subfigure}
    \caption{(a) Energy loss for electrons in argon obtained from NIST ESTAR tables~\cite{bib:NIST_ESTAR}. (b) Energy loss for electrons and positrons in argon. The blue and red curves are obtained employing the formulas presented in ICRU Report 37~\cite{bib:ICRU}, with no density correction applied. The green line is obtained from NIST ESTAR tabulated data~\cite{bib:NIST_ESTAR} (these values are identical to the tabulated quantities in the ICRU report itself, but begin to diverge once density corrections become significant). The magenta line is obtained applying Bethe energy loss to electrons. The bottom section shows the relative difference of plotted curves with respect to the NIST (green) values.}
    \label{fig:electroneloss}
\end{figure}

 Radiative contributions to energy loss from electrons become significant at \unit[10]{MeV}, and are the dominant cause of energy loss by \unit[100]{MeV}. Figure~\ref{fig:electroneloss} (a) shows the energy loss contribution from collisions (ionization) in blue and radiative losses in red as a function of an electron's energy. While ionization losses are continuous over the scale of a few millimeters, radiative contributions are largely stochastic due to the nature of bremsstrahlung and Compton scattering cross-sections, dictating the topological features of EM showers below 1 GeV.

\par Ionization losses for electrons differ slightly from those of heavier particles due to the interaction cross-section with electrons orbiting the nuclei of the target material. Such collisions are described by M{\o}ller scattering, which accounts for the indistinguishability between incoming and target electrons. For positrons, the same interactions are governed by Bhabha scattering. These different interactions lead to a collision stopping power which differs slightly from that of heavier particles described by the Bethe-Bloch formula, and is shown in figure~\ref{fig:electroneloss} (b).

\subsection{Photon Energy Loss}

\par Photons with energies larger than a few MeV lose energy predominantly via $e^+ e^-$ pair production, leading to a cascade that produces EM showers of electrons and photons of successively lower energy. In the few MeV energy range, Incoherent Compton scattering dominates and remains non-negligible up to a few tens of MeV. 
\par Figure~\ref{fig:gammaloss} (a) shows the mean free path $\lambda$ (the inverse of the cross-section) for photons in liquid argon, as a function of the photon energy. In the \unit[$10 \textrm{--} 100$]{MeV} energy range, photons propagate \unit[$20 \textrm{--} 30$]{cm} before undergoing an interaction that leads to energy deposition via electrons. EM showers will thus develop over considerable distances in liquid argon.

\subsection{Stochasticity of EM Showers}

\par The stochastic nature of radiative energy loss causes large event-by-event variations in the topology of EM showers with energies of up to several hundred MeV, where contributions to the energy loss by secondary electrons close to the critical energy (taken here to be the energy at which radiative and collision losses are equal, \unit[39]{MeV}) play a dominant role. The relatively long photon conversion distance and stochasticity of photon production leads to segmented and scattered energy deposit with large gaps which exhibit more variations in topology than higher energy, fully-developed EM showers. 
\par
\begin{figure}[ht]
    \begin{center}
    \begin{subfigure}[b]{0.45\textwidth}
    \centering
    \includegraphics[width=1.00\textwidth]{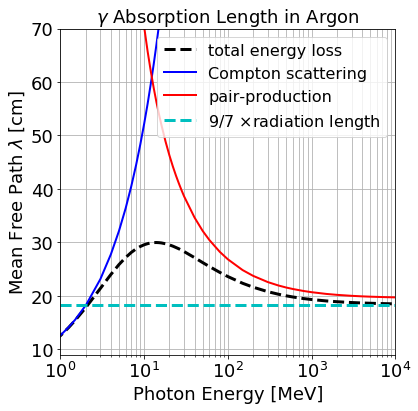}
    \caption{}
    \end{subfigure}
    \begin{subfigure}[b]{0.45\textwidth}
    \centering
    \includegraphics[width=1.00\textwidth]{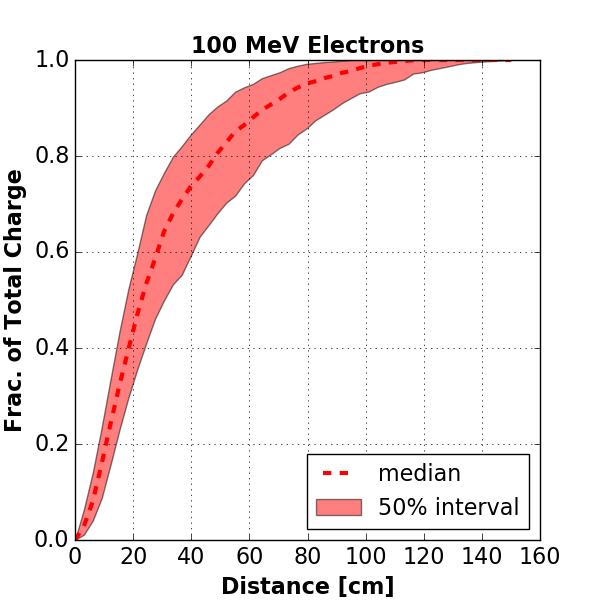}
    \caption{}
    \end{subfigure}
    \caption{(a) Mean free path for photons in liquid argon obtained from NIST XCOM tables~\cite{bib:NIST_XCOM}. In cyan 9/7 the radiation length of \unit[14.1]{cm} is shown, corresponding to the asymptotic mean free path. (b) Energy loss profile for EM showers produced by \unit[100]{MeV} electrons represented by the median fractional energy lost as a function of radial distance from the electron creation point. }
    \label{fig:gammaloss}
    \end{center}
\end{figure}

 At $0.1 \textrm{--} 1$ GeV energies, EM showers deposit their energy over distances of $\sim\!1$ meter, with the shower range logarithmic in energy. Figure~\ref{fig:gammaloss} (b) shows the energy loss profile of EM showers produced by \unit[100]{MeV} electrons as the median fractional energy deposited within a certain radial distance of their starting point. The band denotes the interval encompassing 50\% of all simulated electrons. Its spread is used to estimate the loss of energy resolution caused by the event-by-event variation in energy deposition.

\par
\begin{figure}[ht]
\begin{center}
  \includegraphics[width=0.6\textwidth]{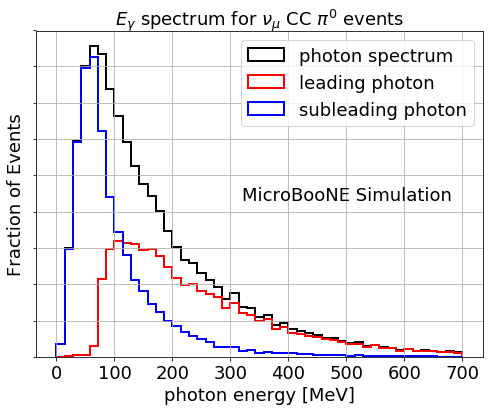}
  \caption{Predicted energy spectrum of photons from the decay of neutral pions obtained from MicroBooNE's $\nu_{\mu} + {\rm Ar} \rightarrow \mu + \pi^0 + X$ simulation assuming the BNB flux. The black line shows the inclusive photon spectrum, subdivided in the two contributions of leading (red) and subleading (blue) photon.}
  \label{fig:gammaspectrum}
  \end{center}
\end{figure}

 Figure~\ref{fig:gammaspectrum} shows the energy distribution of photons produced from the decay of neutral pions from $\nu_{\mu}$ CC interactions as simulated in the MicroBooNE detector. The EM showers being studied in this work mainly populate the \unit[50-200]{MeV} energy range.

\section{Shower Reconstruction}
\label{sec:shrreco}
\par
\begin{figure}[ht]
\begin{center}
  \includegraphics[width=1.0\textwidth]{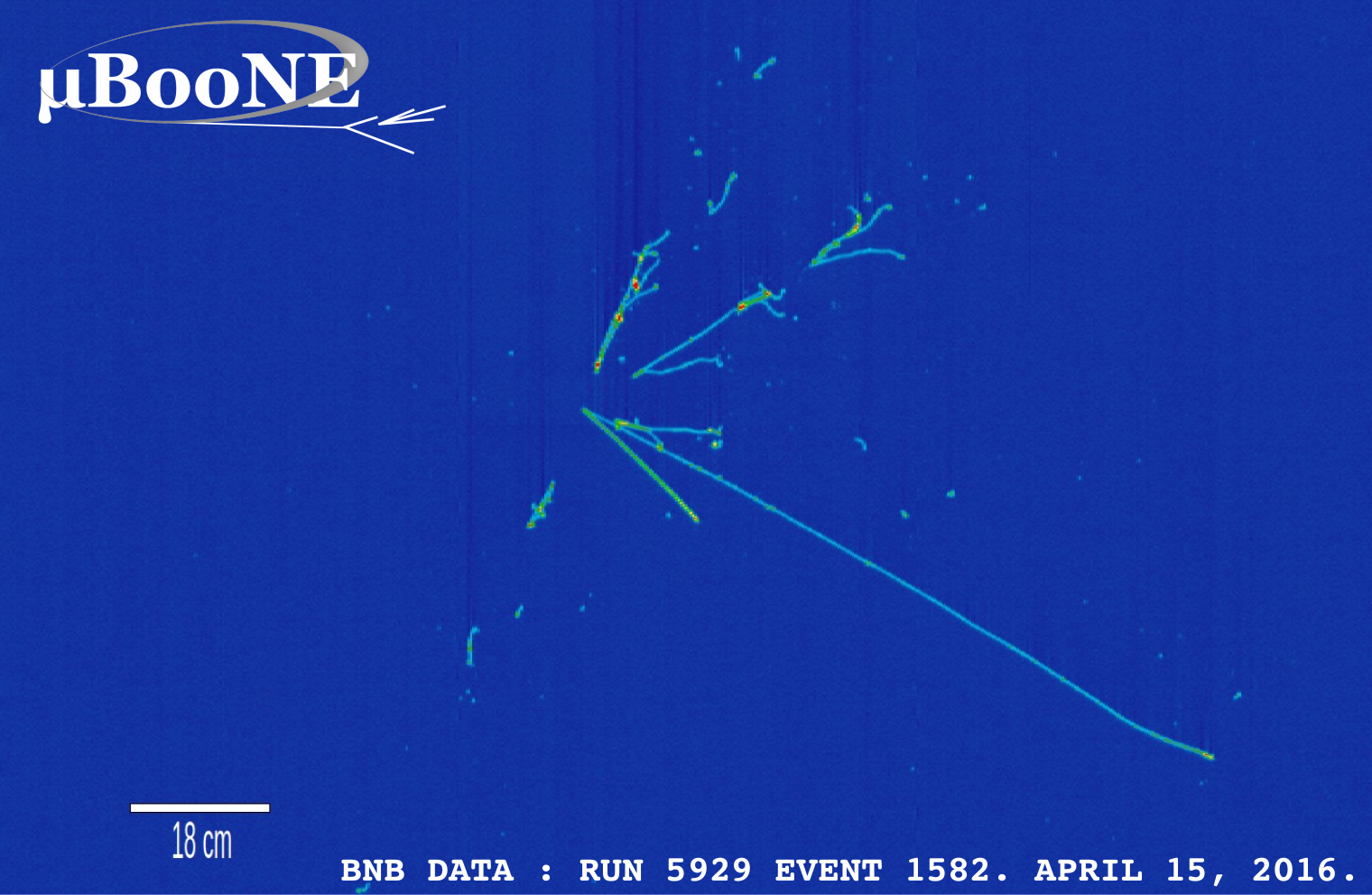}
  \caption{Example of MicroBooNE data event with two $\pi^0$ mesons in the final state. The entire image spans a distance of $\sim$ 1 meter. In this event display the horizontal axis corresponds to the beam direction coordinate, while the vertical to drift time. Color on the image corresponds to the amount of energy deposited. The EM showers in this event, typical for BNB $\pi^0$ events, exhibit a segmented stochastic nature, are segmented, and contain track-like linear segments.}
  \label{fig:2pi0}
  \end{center}
\end{figure}

 There are two main challenges to performing shower reconstruction of EM interactions in a LArTPC: 
\begin{enumerate}
\item It is difficult to separate energy deposition associated with EM showers from that caused by track-like particles. This is a challenging task since showers at the energy of interest for $\pi^0$ reconstruction often appear as scattered track-like segments. Figure~\ref{fig:2pi0} shows an example data event with four EM showers produced in a candidate neutrino interaction.
\item The presence of a high rate of uncorrelated cosmic-ray activity in the event poses a challenge to the energy reconstruction of EM showers, where correctly integrating the energy deposited in the detector is essential. This challenge is particular to a surface detector like MicroBooNE. The significant distance over which showers propagate and the sparsity of energy deposition of low-energy EM showers exacerbate this challenge.	
\end{enumerate}

\par The reconstruction presented in this work is a ``second-iteration'' reconstruction that is performed subsequent to the identification of a sample of candidate neutrino interactions. These interactions are obtained from the neutrino selection described in reference~\cite{bib:ccpi0}. The reconstructed vertex obtained from these candidate interactions is used to guide the $\pi^0$ and $\gamma$ reconstruction.
\par Shower reconstruction is performed in a staged approach. First, hits in the event are classified as shower or track-like by a trained deep-neural network (section~\ref{sec:trkshrseparation}). Shower-like hits are then clustered, employing the neutrino vertex as a guide, searching for radially collinear charge associated with each photon (section~\ref{sec:clustering}). After matching charge from different wire planes (section~\ref{sec:matching}), 3D kinematic properties (start point, direction, energy, and d$E$/d$x$) of showers are reconstructed (section~\ref{sec:showerreco}). To orient the reader in the upcoming decription of the reconstruction in which a specific coordinate system is referenced, a brief description of the MicroBooNE TPC geometry is provided in section~\ref{sec:tpc}.

\subsection{Overview of The MicroBooNE Time Projection Chamber}
\label{sec:tpc}
\par The MicroBooNE TPC is placed on-axis on the BNB and has dimensions of \unit[2.56]{m} in the drift coordinate ($X$), \unit[2.32]{m} in the vertical ($Y$), and \unit[10.36]{m} in the beam direction ($Z$). Ionization charge produced by drifting electrons is detected by recording induced currents on 8,156 wires placed on the anode plane which is oriented vertically, on the $y-z$ plane. Wires are arranged on three wire-planes. The first two planes encountered by the drifting electrons, referred to as induction planes, are at an angle of $+60$ and $-60$ degrees with respect to the vertical direction, and record bipolar signals. The last plane, referred to as the collection plane, has wires aligned vertically and measures uni-polar pulses. For charge deposits by minimally ionizing tracks, induction planes provide signal-to-noise ratios of $10 \textrm{--} 30$, while on the collection plane the range is $30 \textrm{--} 50$~\cite{bib:noise}.
\par The specific orientation of each wire-plane leads to anisotropic charge-detection and reconstruction effects which in turn can cause angular-dependence in detector performance. To illustrate this we describe the case of the collection plane wires, oriented vertically $(0,1,0)$ in detail. Particles moving in the beam direction $(0,0,1)$ are parallel to the collection plane's wire-pitch $(0,0,1)$  and therefore deposit their charge on numerous wires. Particles moving perpendicular to the wire-pitch can either be oriented vertically $(0,1,0)$ or in the drift direction $(1,0,0)$. The former lead to large, isochronous charge deposits which collect on few wires at the same time, while the latter still deposit their charge on very few wires, but this charge is spread in drift, and thus readout time. The orientation-dependent pattern of charge on a wire-plane can lead to angular-dependent reconstruction performance for the identification and reconstruction of charged particle trajectories and EM showers.
\par In addition to angular-dependent charge patterns on the wire-planes, the wire-response itself depends on the orientation of a charged particle's trajectory with respect to the wire-pitch direction. This can lead to biases in the calorimetric estimation of energy loss, particularly for particles moving at large angles with respect to the wire-pitch direction. This causes complications for EM showers, where the energy deposition is contributed by a cascade of electrons and positrons which scatter and move in a broad range of directions. Effects of this angular-dependent charge response impact both the reconstruction of shower d$E$/d$x$, as well as the total calorimetric energy reconstruction.

\subsection{Track-Shower Discrimination and Cosmic Ray Rejection}
\label{sec:trkshrseparation}
\par Identifying energy deposits associated with EM showers is essential to successful shower reconstruction. The detailed nature of LArTPC images and the stochastic variations of EM activity complicate the use of algorithms to achieve this reconstruction.
\par We use an SSNet convolutional neural network~\cite{bib:SSNet} to identify electromagnetic activity in the event. SSNet is an adaptation of the U-ResNet network~\cite{bib:unet,bib:resnet}, which employs deep-learning techniques to identify EM activity in LArTPC neutrino interaction images on a pixel-by-pixel basis. 
The network is trained and assigns a score to each pixel in a 2D wire-versus-time MicroBooNE event image, based on its compatibility with shower- or track-like energy deposits. Scores range from zero to one, with "one" indicating shower-like pixels. Training is performed on Monte Carlo samples of neutrino interactions generated either with the GENIE~\cite{bib:GENIE} neutrino generator or through a single particle generation approach. Training scores are assigned based on the truth-level particle species contributing charge to each pixel in the image. EM-like activity produced by electron and photon showers, as well as Michel electrons and $\delta$ rays, all contribute to the shower-like pixel score. Further details on network training are provided in section IV of reference~\cite{bib:SSNet}.
\par
\begin{figure}[ht]
    \begin{center}
    \includegraphics[width=0.9\textwidth]{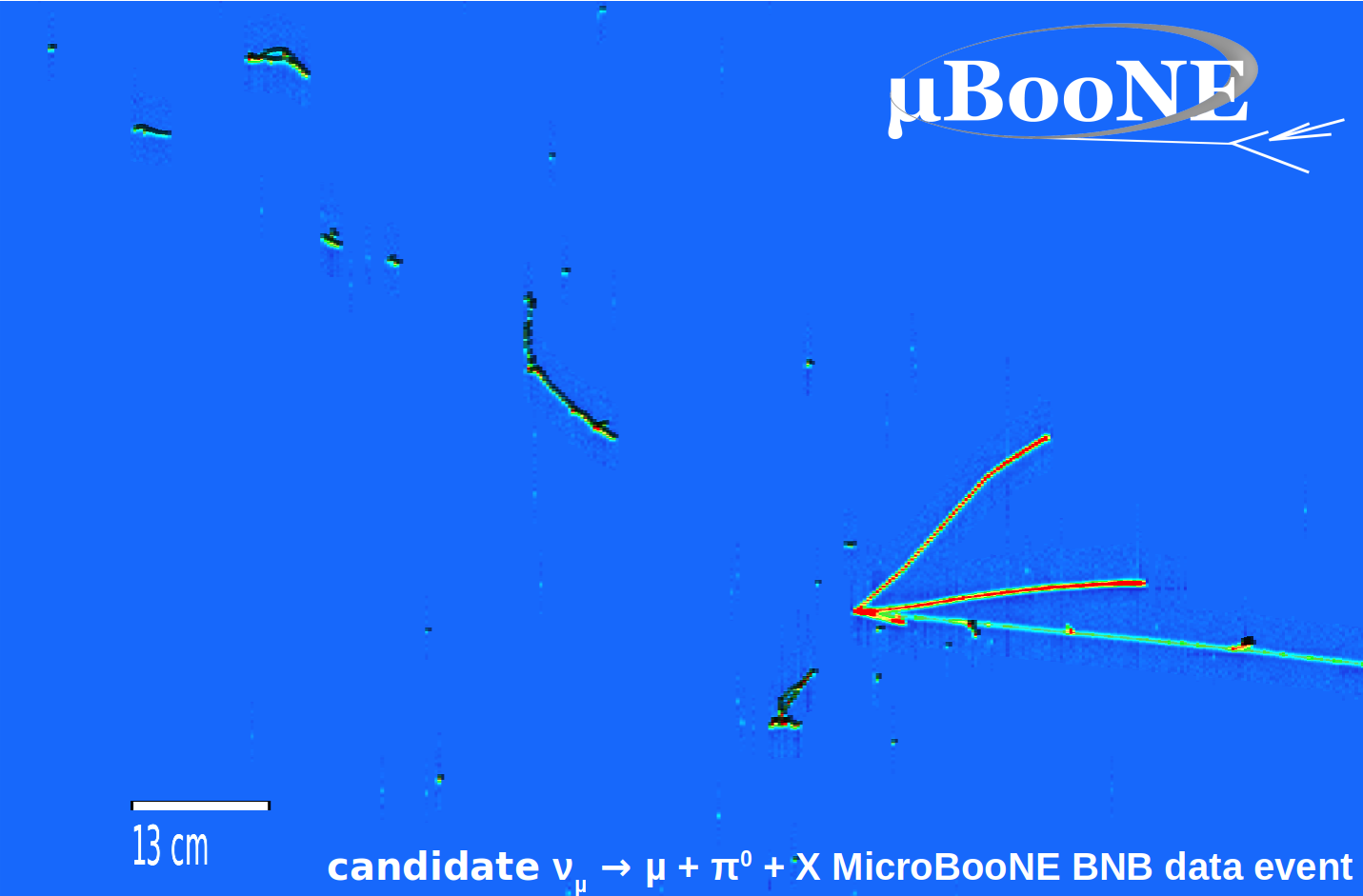}
    \caption{\label{fig:ssnetexample}Candidate $ \nu_{\mu}$ CC $\pi^0$ event from MicroBooNE data. Overlayed in black on the 2D collection-plane image are reconstructed 2D hits with a shower-like score greater than 0.5 as determined by the SSNet network. }
    \end{center}
\end{figure}

 Pixels on each 2D image from the three wire-planes are classified as shower-like if they have a shower score above a given threshold. Hits associated with the 2D coordinates of such pixels are then used in subsequent reconstruction steps as shower-like energy deposition. The SSNet network shows very good performance and data-MC agreement on the collection plane, but less so on induction planes, in which additional noise and signal-processing issues cause noise and track-like pixels to be reconstructed as shower-like. For this reason, collection and induction planes are used in different ways: a high-threshold score of 0.9 is placed on induction-plane hits to reject background hits. Induction-plane hits are used solely to reconstruct the 3D direction and start point of EM showers. Completeness in collecting energy deposition is therefore not a concern on these planes. On the other hand, collection-plane hits are selected as shower-like with a score greater than 0.5. This allows the collection of as much energy as possible deposited by EM showers. Updates in signal processing developed by MicroBooNE~\cite{bib:SP1,bib:SP2} will increase reliance on induction plane information in future analyses.
\paragraph{} An example data event showing a collection-plane view of a candidate $ \nu_{\mu}$ CC $\pi^0$ interaction with overlayed hits from the SSNet pixel-tagging is shown in figure~\ref{fig:ssnetexample}. The ability of the SSNet algorithm to identify shower-like pixels with an accuracy of better then 90\%~\cite{bib:SSNet}, including in cases where EM energy deposition is track-like, as in this example, motivates the choice to implement this tool in our reconstruction approach.

\begin{figure*}[ht] 
\begin{center}
    \begin{subfigure}[b]{0.48\textwidth}
    \centering
    \includegraphics[width=1.00\textwidth]{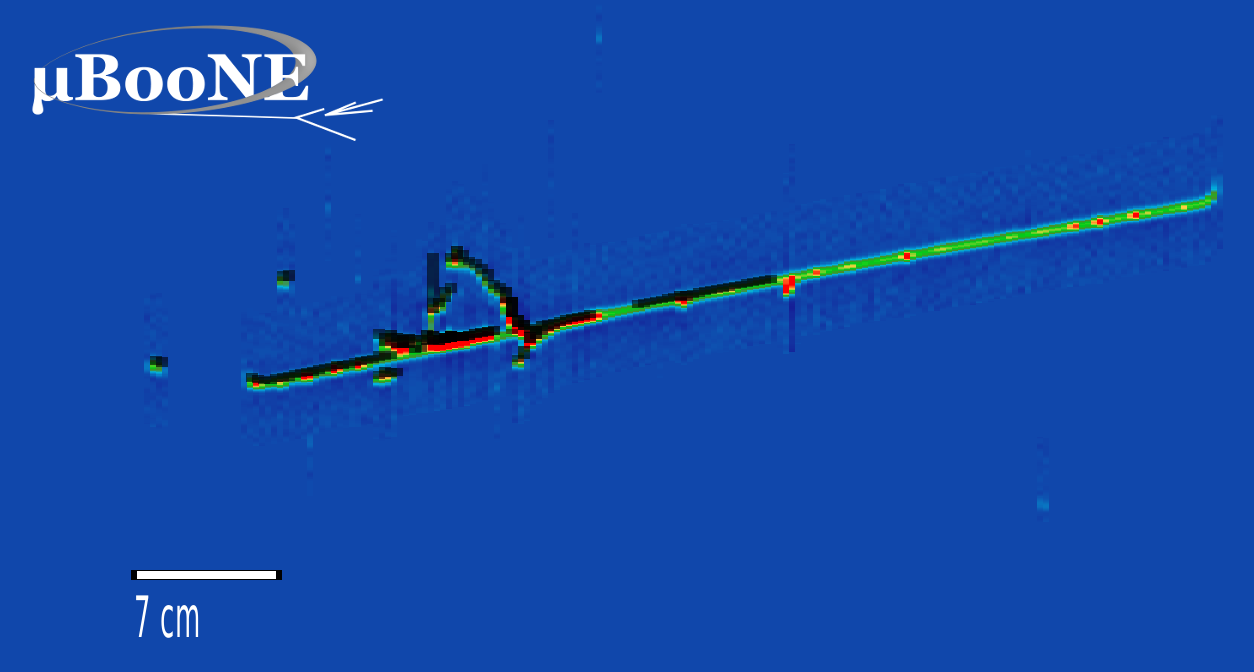}
    \caption{}
    \end{subfigure}
    \begin{subfigure}[b]{0.48\textwidth}
    \centering
    \includegraphics[width=1.00\textwidth]{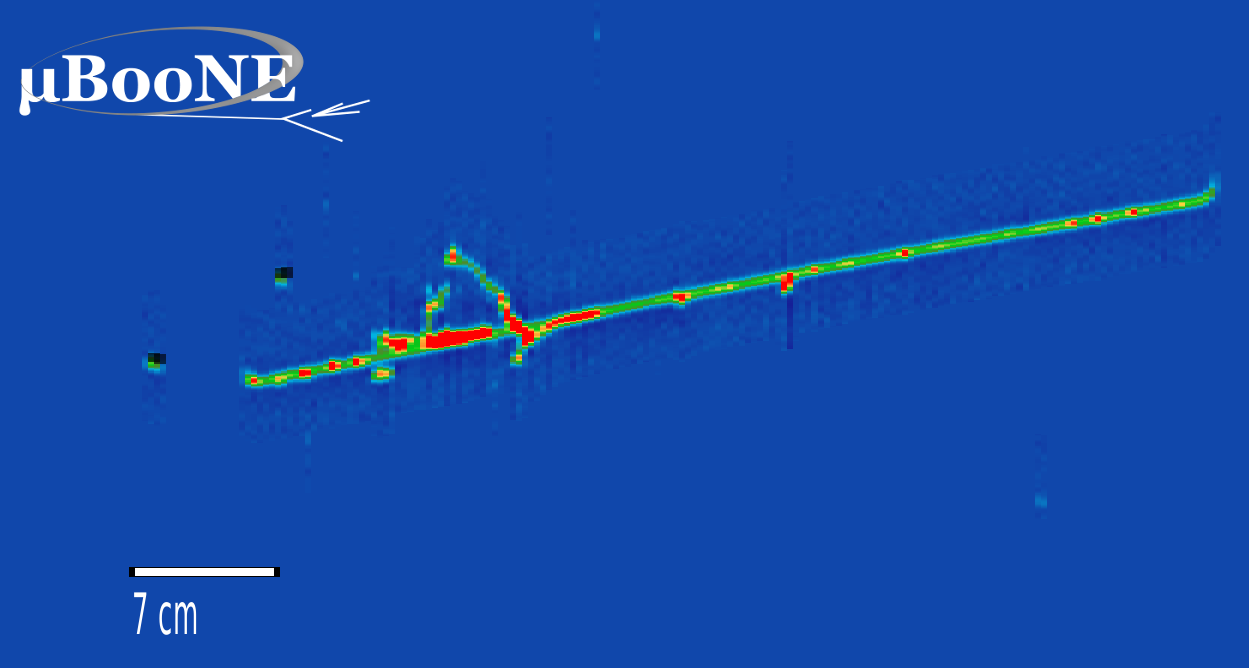}
    \caption{}
    \end{subfigure}
\caption{\label{fig:ssnetbkgd}MicroBooNE data event showing SSNet shower-like pixels for a cosmic-ray. Black hits denote EM-like pixels as identified by SSNet. (a) Before cosmic-hit removal. (b) After cosmic-hit removal.}
\end{center}
\end{figure*}

\paragraph{} The SSNet network is trained to discriminate between energy deposited by electrons and photons, and that produced by track-like particles (protons, pions, muons). It therefore associates EM activity correlated with muons, such as $\delta$ rays and bremsstrahlung showers, as shower-like. In addition, SSNet pixel tagging often associates energy deposited by muons in proximity to such correlated EM activity to a shower. These types of interactions are a background to the $\pi^0$ reconstruction. Figure~\ref{fig:ssnetbkgd} shows an example cosmic-ray muon from data with shower-like hits in proximity to EM activity correlated with the muon. Because of the cosmic-ray background in MicroBooNE data and the similar energy deposited by cosmic rays and $\gamma$ showers, we target the removal of these backgrounds with specific selection criteria. The PANDORA pattern-recognition cosmic-ray muon reconstruction (section 4.1 of reference~\cite{bib:pandora}) is used to identify cosmic muon tracks in 3D. Those longer than \unit[50]{cm} are selected. If such tracks have an Impact Parameter ($IP$) with respect to the neutrino vertex greater than \unit[10]{cm}, all shower-like hits associated either to the track itself or to any correlated $\delta$ rays are removed. Remaining shower-like hits from the cosmic muon of figure~\ref{fig:ssnetbkgd} after these cuts are applied are shown in the image on the right.

\subsection{Charge Clustering}
\label{sec:clustering}

\begin{figure}[ht]
    \centering
    \begin{subfigure}[b]{0.32\textwidth}
    \centering
    \includegraphics[width=1.00\textwidth]{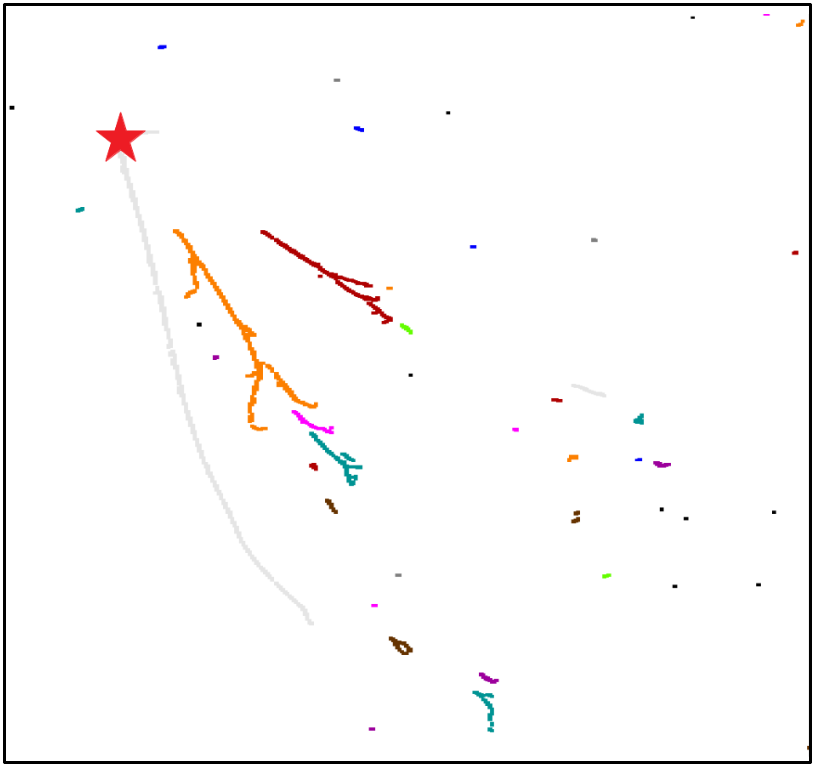}
    \caption{}
    \end{subfigure}
    \begin{subfigure}[b]{0.32\textwidth}
    \centering
    \includegraphics[width=1.00\textwidth]{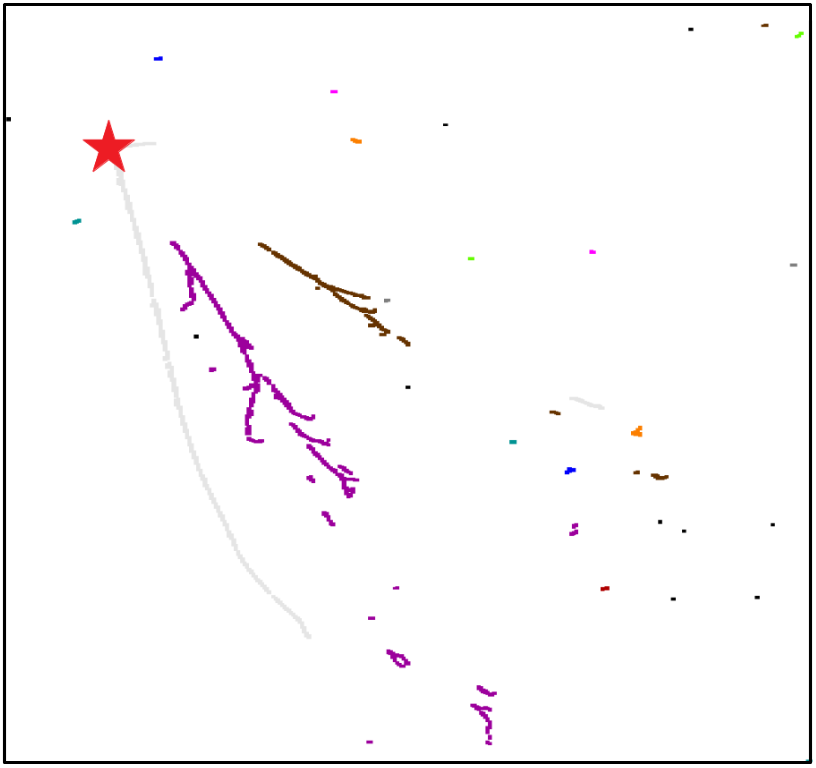}
    \caption{}
    \end{subfigure}
    \begin{subfigure}[b]{0.32\textwidth}
    \centering
    \includegraphics[width=1.00\textwidth]{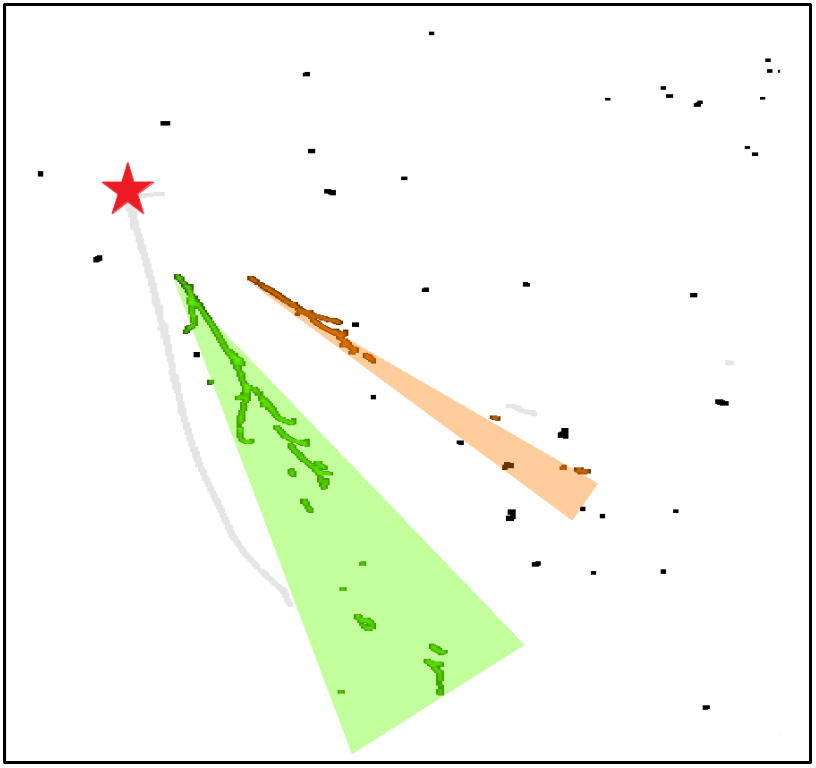}
    \caption{}
    \end{subfigure}
    \caption{Example event display from a simulated $\nu_{\mu} + {\rm Ar} \rightarrow \mu + \pi^0$ event in MicroBooNE. Track-like pixels as identified by SSNet are masked out (grey), and the reconstructed vertex is labeled with a red star. (a) Individual segments of the photon showers, clustered by proximity, are labeled by different colors. (b) The photon showers as identified by the clustering algorithms are labeled in magenta and brown. (c) The reconstructed showers, shown as green and orange cones, are overlayed on hits with corresponding colors associated to the two showers on the collection plane. The orange shower shows an example of under-clustering, with some hits not correctly associated to the reconstructed photon.}
    \label{fig:recoshowerexample}
\end{figure}

\paragraph{} This reconstruction stage is tasked with grouping the reconstructed hits on each plane into clusters, one for each of the two $\gamma$s produced by the $\pi^0$. This is done through a series of algorithms, and precedes the full 3D shower reconstruction. Figure~\ref{fig:recoshowerexample} shows an example event through the clustering stages (a, b) and final 3D reconstruction (c).

Hits identified as shower-like in the track-shower separation stage (section~\ref{sec:trkshrseparation}) are clustered via a proximity algorithm into independent and contiguous charge segments (see figure~\ref{fig:recoshowerexample} (a)). These proximity-based clusters are used as the input to the $\gamma$ clustering stage for which two algorithms are applied.


\subsubsection{Polar Coordinate Merging}
\par This algorithm is a specific implementation of a widely used class of cone clustering algorithms. Hit coordinates in wire and time are converted to polar coordinates with the neutrino vertex as the origin. The radial and angular correlations of individual energy deposits are used to guide their merging into a single photon shower. For each cluster we compute an \textit{angle} and \textit{angle-span}, defined, respectively, as the charge-weighted direction of the hits in the cluster, and the range of angles encompassing all hits in it. A start and end point are also reconstructed, corresponding to the ($r$,$\theta$) coordinates of, respectively, the hit closest and furthest away from the vertex. The radial distance between these two points is referred to as the \textit{cluster length}. In this context, the qualifiers \emph{upstream} and \emph{downstream} for two clusters are used to denote the one closest and the one furthest from the neutrino interaction vertex, respectively. Using these quantities, the showers are merged if all of the following criteria are met:
\begin{enumerate}
\item The upstream shower has more charge associated to it than the downstream one.
\item The downstream cluster reconstructed angle $\theta$ is within the angle-span of the larger upstream one. 
\item The distance between the two clusters, measured as the radial separation between the upstream end-point and the downstream start-point, is smaller than the total length of the larger upstream photon-cluster.
\end{enumerate}
This procedure is repeated recursively on each plane separately until an iteration is reached in which no further clusters are merged.

\subsubsection{Vertex-Aligned Merging}
\par A second algorithm aims at clustering photon-clusters under the assumption that two $\gamma$ EM showers from a $\pi^0$ decay are present in the event. The first action taken is identifying, in each plane, the two clusters with the largest amount of charge. These two clusters must be separated in polar angle $\theta$ by at least 15 degrees. Once these two clusters are identified, all remaining clusters in the event are scanned to determine if they should be merged with a $\gamma$ cluster. A smaller cluster is merged if the following conditions are both met:
\begin{enumerate}
\item Its polar angle is within 12 degrees of the closest large $\gamma$ cluster, but more than 15 degrees away from the second large $\gamma$ cluster.
\item The radial distance between its start point and the large gamma cluster end point is less than three times the radial length of the large $\gamma$ cluster.
\end{enumerate}
This procedure is repeated recursively on each plane until an iteration is reached in which no more clusters are merged. As a reminder, the clustering techniques employed are purposely conservative in order to avoid over-clustering EM activity associated with uncorrelated cosmic-rays in the event.

\subsection{Clustering Inefficiencies}
\par
\begin{figure}[h] 
\centering
\includegraphics[width=0.6\textwidth]{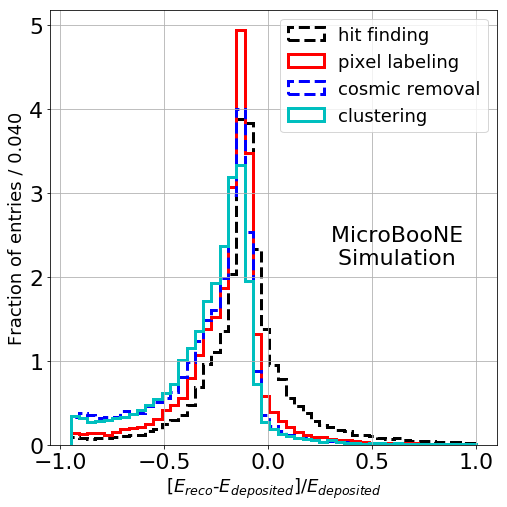}
\caption{Fractional energy resolution measuring the impact of clustering and other charge-collection inefficiencies, evaluated on a sample of 10s to 100s of MeV photon EM showers from simulated CC $\pi^0$ neutrino interactions using collection-plane reconstructed charge.}
\label{fig:ineff}
\end{figure}

 Inevitably, each step in the reconstruction can lead to inefficiencies in recovering the full energy deposited by EM showers. We examine the inefficiencies here, studying in the simulation the deficit between the energy recovered at each step in the reconstruction versus the true photon energy. The metric we utilize is the fractional energy difference $[E_{\rm reco}-E_{\rm deposited}]/E_{\rm deposited}$ with $E_{\rm deposited}$ the photon energy deposited in the TPC active volume from simulation, and $E_{\rm reco}$ the energy associated with surviving hits on the collection plane. Figure~\ref{fig:ineff} reports this study for the four stages in our reconstruction, evaluated on a sample of photons from simulated BNB $\nu_{\mu}$ CC $\pi^0$ neutrino interactions with overlayed simulated CORSIKA~\cite{bib:CORSIKA} cosmic-ray particles. 

\par Hit finding, in black in the figure, encapsulates the effect of identifying signals produced by energy deposits on the wires and reconstructing hits associated with them. The distribution at this stage is largely symmetric around zero with entries above zero caused by energy response smearing in the reconstruction. Nonetheless, a negative offset of 10\% in the peak of the distribution is present, associated with thresholding effects which impact EM showers due to the significant number of energy depositions of less than a few hundred keV. In simulation, we find that the hit reconstruction threshold on the collection plane is approximately \unit[300]{keV}. Pixel-labeling, in red in the figure, begins to significantly skew the distribution, adding to the bias and to the low-end tail in the distribution. The bias intensifies further after the cosmic-removal stage, in blue in the figure, particularly with a significant broadening of the low-end tail due to events for which many of the photon hits are removed due to their accidental proximity to a cosmic-ray muon. Finally, the clustering stage, in cyan in the figure, exhibits an additional residual inefficiency. In moving from hit finding to clustering, the purity of the collected photon showers is increasing. The skewed and biased nature of the distribution of figure~\ref{fig:ineff} will impact energy reconstruction for EM showers, as will be discussed in detail in Section~\ref{sec:ereco}. The purity of clustered charge is also evaluated and found to be on average 82\%.

\subsection{Cluster Matching}
\label{sec:matching}
\par After clusters have been merged on each plane separately, a cross-plane matching algorithm is applied to associate clusters belonging to the same $\gamma$. Due to wire-coverage in the TPC, and the overall difficulty of finding a photon which is well-clustered on all three planes, we require that matching be performed only between pairs of planes. We further demand that one of the two clusters must be associated to the collection plane, as this will provide us with the best information with which to perform calorimetry and measure the energy of the $\gamma$. Finally, only clusters which have at least 10 reconstructed hits will be considered. The matching algorithm applied calculates the overlap in time of pairs of clusters from different wire-planes and assigns a score based on this overlap. Cluster-pairs with the highest score are then merged. The figure of merit devised as the overlap score is denoted \emph{IoU}, for Intersection over Union, and is defined as the time-interval common to the two clusters over the union of the two clusters' time-spans. A minimum \emph{IoU} of 0.25 is required to match two clusters, and clusters are matched in order of their score: if a cluster has a match with two or more other clusters, the pair with the largest score is associated as belonging to the same $\gamma$. 

\subsection{3D Shower Reconstruction}
\label{sec:showerreco}
\par  Finally, pairs of clusters on two planes are used to reconstruct 3D showers. Specifically, this reconstruction stage aims to measure the $\gamma$'s 3D start point and direction, as well as its energy and dE/dx. The reconstruction is modularized in a series of algorithms, which are described below.

\par The 3D direction of a shower is reconstructed by geometrically correlating the two 2D directions obtained on the pair of matched clusters. The 2D direction on each plane is calculated by measuring the charge-weighted 2D direction of hits in the cluster with respect to the neutrino vertex location. An accuracy in shower direction of 3 degrees is obtained from simulation studies.

\par The 3D start point is reconstructed by taking the reconstructed start-point in 2D associated with the collection-plane cluster, and projecting it onto the reconstructed 3D direction to recover the missing $Y$ (vertical) coordinate. The $X$ (drift) and $Z$ (beam) coordinates are reconstructed from the time-tick and wire associated with the start-point on the collection-plane.
\par The shower energy is reconstructed calorimetrically, by integrating all the charge associated with collection-plane hits belonging to the $\gamma$ shower, and converting this quantity to MeV by accounting for the following factors:
\begin{itemize}
\item An electronics gain obtained from calibration of the d$E$/d$x$ of minimally ionizing muons that stop in the detector.
\item A work function for ionization in liquid argon of 23.6 eV/$e^-$~\cite{bib:W}.
\item An effective recombination factor $R$, obtained from studies presented in Section~\ref{sec:ereco}.
\end{itemize}
The energy reconstruction of EM showers, methodology for correcting for energy biases, and energy resolution studies, are presented in detail in section~\ref{sec:ereco}.

\par The energy deposited by an EM shower in the first few centimeters of propagation can help distinguish electrons from photons, given sufficient calorimetric and spatial resolution. For each shower, a dE/dx quantity is calculated by considering hits within a four centimeter radius of the shower starting-point. Charge from these hits is integrated in 3 $mm$ segments extending radially in the shower direction. The median value of non-empty segments is then chosen as the shower dE/dx. Results for reconstructed $\gamma$ candidates are presented in Section~\ref{sec:dedxpi0}.

\par A quality cut is applied to ensure that reconstructed showers are truly associated with neutrino-induced photons. We compare the reconstructed 3D shower direction, projected on the collection-plane, and the charge-weighted vector sum computed from the neutrino vertex to collection-plane 2D hits. If the angle between these vectors is larger than 25 degrees, the shower candidate is rejected. 

\par To improve the clustering efficiency and thus the energy reconstruction of EM showers, a second clustering stage is applied once 3D showers are reconstructed. The advantage of applying this second-pass clustering step is that, given a candidate 3D shower, it is easier to identify charge spatially correlated with the shower direction. This charge may have been missed at an earlier reconstruction stage due to a conservative clustering approach which purposely attempts to avoid including accidental charge in the shower. Clusters are merged into an already existing shower if they overlap a 2D projected cone \unit[150]{cm} long, with an opening angle of 30 degrees and with its apex at the shower start point. If the overlapping cluster contains more than eight hits, two further requirements are imposed. The direction of the 2D hits of the shower and photon cluster to be merged, calculated via a linear regression to the hit coordinates, must agree to within 30 degrees. In addition, the cluster to be merged cannot cross the projected cone on more than one boundary.

\subsection{Shower Reconstruction Performance}
\label{sec:showerrecoperformance}
\par
\begin{figure}[h] 
  \begin{center}
    \begin{subfigure}[b]{0.46\textwidth}
    \centering
    \includegraphics[width=1.00\textwidth]{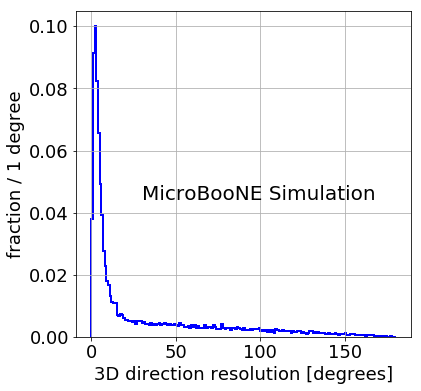}
    \caption{}
    \end{subfigure}
    \begin{subfigure}[b]{0.44\textwidth}
    \centering
    \includegraphics[width=1.00\textwidth]{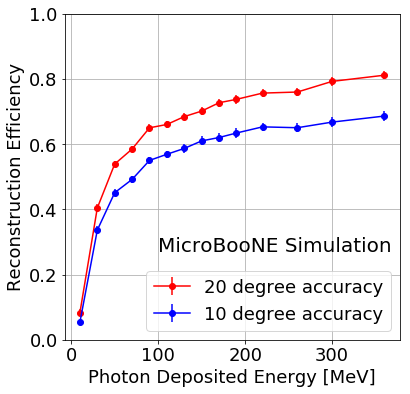}
    \caption{}
    \end{subfigure}
  \caption{(a) 3D angular resolution for reconstructed EM showers. (b) Reconstruction efficiency of $\gamma$ EM showers from simulated CC $\pi^0$ interactions as a function of the photon's energy deposited in the detector. The efficiency calculation has as denominator all true $\gamma$ photons produced in the active volume from simulated neutrino $\pi^0$ decays. The numerator includes all such true photons with an associated reconstructed EM shower reconstructed within 10 degrees of the true direction. }
  \label{fig:shrrecoeff:energy}
  \end{center}
\end{figure}

 The performance of the reconstruction is evaluated on simulated $\nu_{\mu}$ CC interactions with final state $\pi^0$s and overlayed simulated cosmic ray interactions. The peak in the angular resolution plot is 2.7 degrees, with 23\% (60\%, 70\%) of simulated photons associated to a reconstructed EM shower within 3 (10, 20) degrees (see figure~\ref{fig:shrrecoeff:energy} (a)). The shower reconstruction efficiency as a function of true $\gamma$ deposited energy in the TPC is shown in figure~\ref{fig:shrrecoeff:energy} (b). We note that we achieve a shower reconstruction efficiency of at least 60\% for showers with more than \unit[100]{MeV} of deposited energy and that reconstruct to within 10 degrees of the true shower direction. The efficiency drops for energies below \unit[100]{MeV}, mostly due to the challenge of identifying low-energy EM showers due to their topological features. The efficiency is found to depend on shower direction as well, with a decrease in efficiency for showers propagating in the drift and vertical directions, perpendicular and parallel to the collection-plane wire directions respectively. This is because the shower projection on the collection plane is particularly challenging to reconstruct if the charge is collected on only a few wires. Energy reconstruction performance studies are presented in detail in section~\ref{sec:ereco}.

\section{$\pi^0$ Event Selection}
\label{sec:pi0selection}
\par Candidate $\pi^0$ events are selected by applying the following criteria:
\begin{enumerate}
\item Events must have two or more reconstructed showers. In cases with more than two EM showers, only the two highest energy showers are considered. These two showers are the $\gamma$ candidates.
\item Each $\gamma$ candidate must have more than \unit[30]{MeV} of reconstructed energy.
\item The two EM showers must have an opening angle between them greater than 20 degrees. Small opening angles are often indicative of events in which a single EM shower was split into two reconstructed objects. A 20 degree requirement corresponds to excluding $\pi^0$ momenta greater than \unit[700]{MeV}, which is far in the tail of the distribution of $\pi^0$ momenta expected from BNB neutrino interactions.
\end{enumerate}
\par
\begin{figure}[h] 
  \begin{center}
  \includegraphics[width=0.6\textwidth]{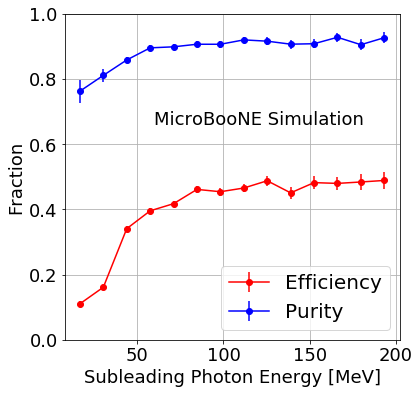}
  \caption{Efficiency and purity of the $\pi^0$ selection evaluated on simulated $\nu_{\mu}$ CC events with a $\pi^0$ in the final state. The efficiency is measured relative to the $\nu_{\mu}$ CC pre-selection. Both efficiency and purity are measured as a function of the energy of the least energetic of the two $\pi^0$ decay photons.}
  \label{fig:pi0_effpur}
  \end{center}
\end{figure}

 The efficiency and purity of this selection, measured with respect to the underlying efficiency of identifying neutrino candidate events, is shown in figure~\ref{fig:pi0_effpur}. The mis-identification rate for neutrino interactions in simulation is found to be 0.75\%, which, due to the abundance of non-$\pi^0$ CC interactions, leads to a purity of approximately 80\%. Backgrounds are dominated by categories in which EM activity in the event originates either from charge-exchange interactions induced by a final-state charged pion which exits the target nucleus ($\pi^{+} \rightarrow \pi^0 + p$) or events where cosmic-ray EM activity near the neutrino vertex is mis-associated. The $\pi^0$ selection efficiency is heavily dependent on the energy of the sub-leading $\gamma$ shower, and saturates at approximately 50\% for events with a subleading shower energy greater than \unit[50]{MeV}.

\begin{figure} 
    \begin{center}
    \includegraphics[width=0.9\textwidth]{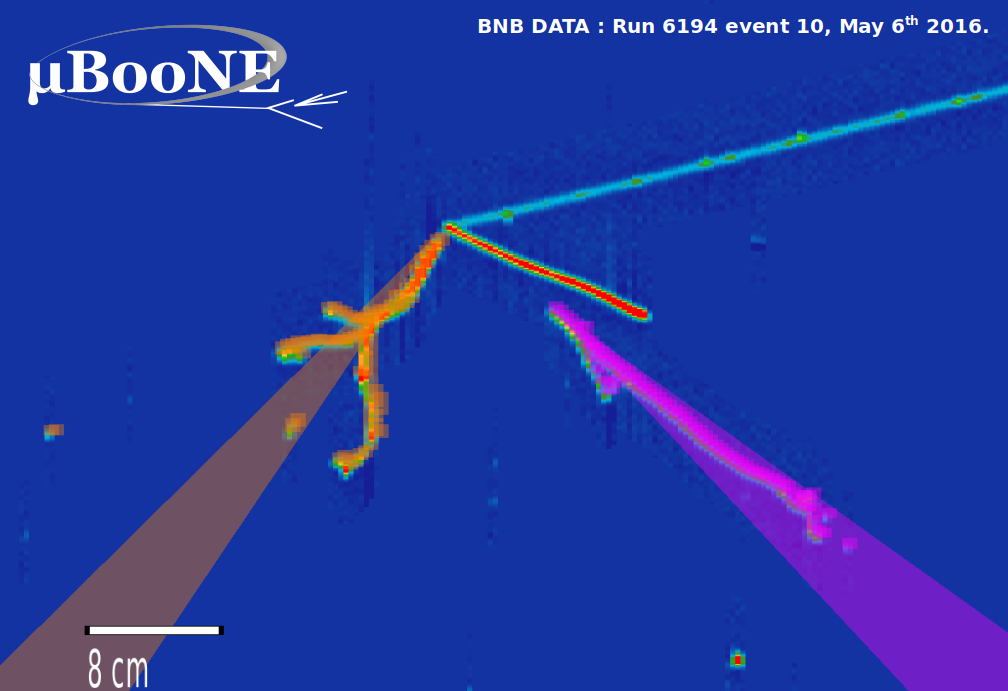}
    \caption{Example $\nu_{\mu}$ CC $\pi^0$ candidate event from MicroBooNE data. Reconstructed showers are overlayed on the event in orange and purple. The long track exiting on the right hand side of the image is the candidate muon, while the short track in red is likely a proton.}
    \label{fig:pi0protonevd}
    \end{center}
\end{figure}

When applied to $1.6\times10^{20}$ protons on target (POT) of BNB data, collected from February to July 2016, the selection leads to the identification of $440 \pm 21$ candidate $\nu_{\mu}$ CC $\pi^0$ events (of which 88 expected background), one of which is shown in figure~\ref{fig:pi0protonevd}.

\section{Energy Reconstruction}
\label{sec:ereco}
\par This section presents the energy reconstruction performed for $\gamma$-induced EM showers which includes a data-driven validation of the calorimetric energy reconstruction on muons, an evaluation of energy reconstruction biases, and corrections from simulation, as well as a profiling of the energy reconstruction performance.

\subsection{Calorimetric Energy Reconstruction}
\par Shower energy reconstruction via calorimetry is performed by integrating the charge recorded by TPC wires associated with EM activity and recovering a calibrated MeV energy scale. Doing so requires accounting for detector effects such as charge loss due to electron attenuation and ion recombination. A calibration to convert pulse amplitudes collected on the wires to drifting charge (in units of number of electrons) is obtained using a sample of stopping muons for which energy loss profiles are known~\cite{bib:PDGmuons}. This calibration procedure is described in reference~\cite{bib:calib}. Charge quenching due to recombination is modeled using ArgoNeuT's modified box parametrization~\cite{bib:argoneutrecomb}, applied for MicroBooNE's electric field of \unit[273]{V/cm}. For this work, attenuation due to electron lifetime and space-charge effects are not corrected for, leading to $2\textrm{--}3$\% smearing in the energy resolution in the bulk of the detector volume. Additional details on the absolute energy scale calibration applied for muons in MicroBooNE can be found in reference~\cite{bib:calib}.
\par Calorimetric energy reconstruction can be validated on stopping muons for which a range-based energy measurement is also obtainable. This data-driven comparison, performed using a sample of tagged stopping muons, shows agreement at the 3\% level. The calorimetric energy reconstruction procedure applied to muons for this sample is identical to that applied to photons, giving confidence in the energy-scale calibration. While the same ion recombination model is used, the implementation of corrections to account for this effect is different for showers, and discussed in detail in the next section.
\par We assess an uncertainty in the energy scale calibration for this work of 3\% for charge deposited collinear to the collection-plane wire-pitch direction, noting that additional angular-dependent biases can impact the energy reconstruction of showers in particular.

 \subsection{Ion Recombination for EM Showers}
 
 \par
\begin{figure*}[ht]
\centering
    \begin{subfigure}[b]{0.53\textwidth}
    \centering
    \includegraphics[width=1.00\textwidth]{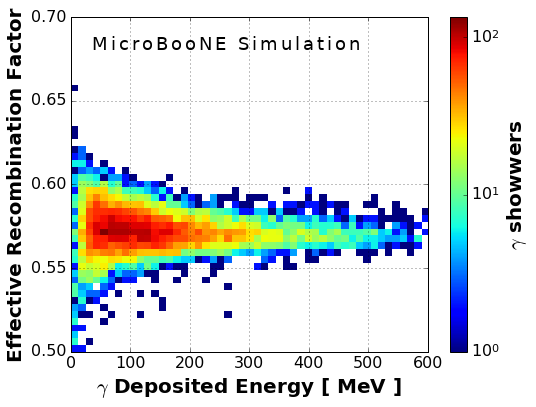}
    \caption{}
    \end{subfigure}
    \begin{subfigure}[b]{0.43\textwidth}
    \centering
    \includegraphics[width=1.00\textwidth]{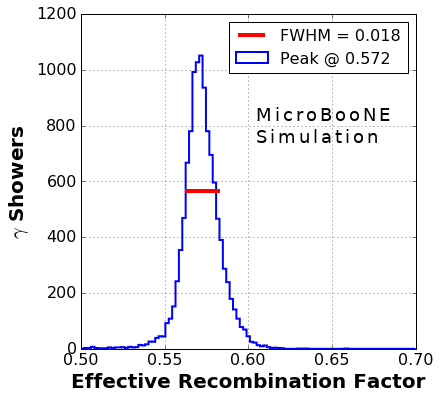}
    \caption{}
    \end{subfigure}
\caption{(a) Effective recombination factor for simulated $\gamma$ showers vs. true photon energy. (b) Distribution of the effective recombination factor for all photons. The full-width at half maximum (FWHM) is an estimate of the smearing in energy response caused by the use of an effective recombination correction.}
\label{fig:recomb_shr_vs_e}
\end{figure*}

 Ionization electrons can recombine with positive argon ions produced concurrently in a charged particle's energy loss. In terms of magnitude, recombination is the largest physics effect that impacts energy collection, suppressing almost half of the energy deposited in the detector. When correcting for ion recombination, the observables to take into account are the local electric field and the local energy deposition density, which can be related to the observable d$E$/d$x$. For particles which deposit energy at different rates along their path, such as stopping muons or protons, accounting for the significantly varying recombination factor at different steps in a particle's path is essential to recover the correct calorimetric energy measurement. In the case of EM activity, which has a much flatter energy loss rate over a wide energy range,  this correction is much more uniform. In addition, calculating an accurate path length $dx$ necessary to recover the correct recombination factor step-by-step is made difficult by the fact that reconstructing the 3D direction of EM energy deposition hit-by-hit is very challenging for the bulk of the shower. We therefore decide to implement an \emph{effective recombination} correction applied to the total measured shower charge. The effective recombination factor, $R_{\rm effective}$, accounts for the global impact of charge quenching on a given EM shower, and is defined in equation~\ref{eq:effrecomb} as the fraction of charge surviving after recombination, where $Q_{\rm visible}$ is the total number of drifting electrons released after quenching, $E_{\rm deposited}$ the total energy deposited by the EM shower, and $W_{\rm ion}$ the work function of argon. and is meant to account for the global impact of charge quenching on a given EM shower. 
 \begin{equation}
 \label{eq:effrecomb}
     R_{\rm effective} = \frac {Q_{\rm visible} \, [e^{-}] \times W_{\rm ion} \, [{\rm MeV}/e^{-}]}{E_{\rm deposited} \, [{\rm MeV}]}
 \end{equation}
 This quantity is plotted for a sample of $\gamma$ showers from simulated muon neutrino CC $\pi^0$ events as a function of the photon energy in figure~\ref{fig:recomb_shr_vs_e} (a). The same distribution is plotted in figure~\ref{fig:recomb_shr_vs_e} (b) and shows, collapsed in one dimension, a peak value of 0.572 with a spread of 0.018. We take this as the spread in energy resolution introduced by applying a constant recombination factor for measuring the reconstructed shower energy. This term is smaller then other contributions to energy resolution, which are dominated by inefficiencies in the shower charge integration. 

\subsection{Energy Biases and Corrections}
\label{sec:ereco:ecorr}
\par
\begin{figure}[ht] 
  \centering
  \includegraphics[width=0.75\textwidth]{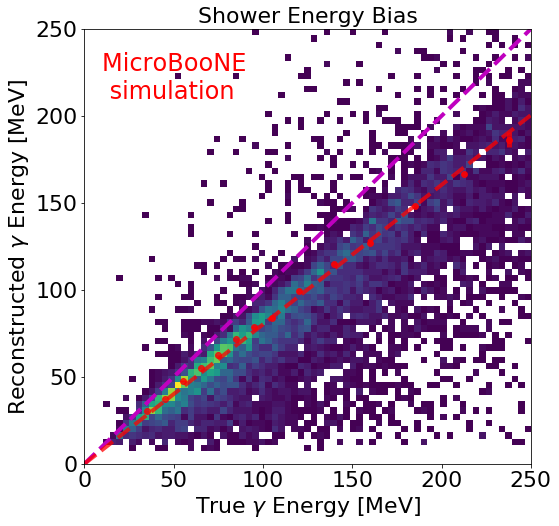}
  \caption{Reconstructed vs. true energy for reconstructed $\gamma$ EM showers from simulated $\pi^0$ decays. The magenta line represents $E_{\rm reco} = E_{\rm true}$. The red points are the medians of the Gaussian plus exponential tail distributions fitted in bins of true energy. The red line (slope $0.802$, intercept $0$) is the result of a linear fit, constrainted to pass through the origin, to the points. The magenta line deontes the diagonal $y = x$.}
  \label{fig:reco_vs_true_pi0}
\end{figure}

 Thus far we have validated the procedure for calorimetric energy reconstruction on stopping muons, and established that an effective recombination correction can adequately account, with minimal smearing, for the effect of charge loss due to recombination. We next evaluate the performance of this energy reconstruction procedure on EM showers in simulation. Figure~\ref{fig:reco_vs_true_pi0} compares the reconstructed to true $\gamma$ energy for $\gamma$ showerss from simulated $\nu_{\mu}$ CC $\pi^0$ interactions. Unlike for muons, a significant bias is observed. This bias is introduced by two main effects: charge falling below the threshold necessary to identify and reconstruct a hit (\textit{thresholding}) and inefficiencies due to hits not correctly labeled as shower-like, or not accounted for in the 2D reconstruction of shower clusters (\textit{under-clustering}). The intrinsically lossy nature of these processes leads to an under-estimation of the total shower energy. These biases must be accounted for in order to recover the correct energy scale and reconstruct the kinematics of the photons and hence of the $\pi^0s$ which produced them.

\par
\begin{figure*}[ht] 
  \centering
    \begin{subfigure}[b]{0.32\textwidth}
    \centering
    \includegraphics[width=1.00\textwidth]{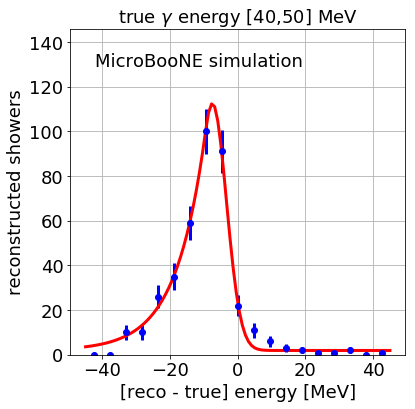}
    \caption{}
    \end{subfigure}
    \begin{subfigure}[b]{0.32\textwidth}
    \centering
    \includegraphics[width=1.00\textwidth]{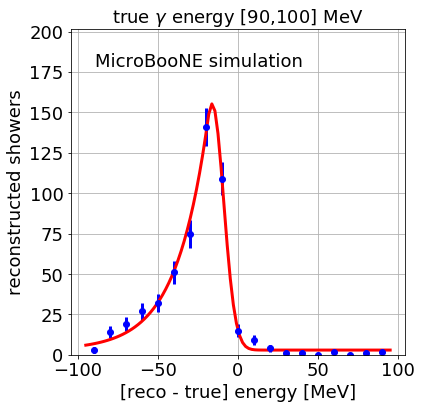}
    \caption{}
    \end{subfigure}
    \begin{subfigure}[b]{0.32\textwidth}
    \centering
    \includegraphics[width=1.00\textwidth]{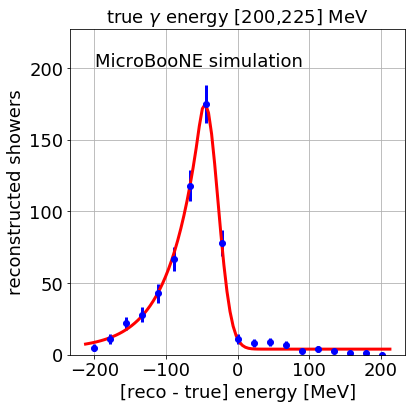}
    \caption{}
    \end{subfigure}
  \caption{Energy resolution for reconstructed $\gamma$ showers from simulated $\pi^0s$. The blue points indicate the measured energy difference. The red curves denote the fitted Gaussian plus one-sided exponential functions used to model energy smearing. The importance of modeling lossy contributions from clustering inefficiencies and thresholding effects can be clearly seen from the substantial negative energy difference of these distributions.}
  \label{fig:reco_vs_true_040050}
\end{figure*}

 In order to quantify the bias observed, and be able to correct for it, we extract, in bins of true energy, the fractional energy resolution defined as $(E_{\rm reco}-E_{\rm true})/E_{\rm true}$, and fit each distribution to a Gaussian plus a low-tail exponential. This choice is motivated by the interest in modeling the lossy impact of clustering and thresholding on energy reconstruction. Examples of such distributions and the resulting fits are shown in figure~\ref{fig:reco_vs_true_040050} for three energy ranges. The mean of the fitted Gaussian is taken as an estimate of the most-probable energy bias for each true energy bin. The bias is found to range between 10\% and 20\%, depending on the energy bin. 

The measured energy bias is fit to a straight line constrained to pass through the origin and its slope is used to compute an energy-independent correction factor which aims to account for the bias. Figure~\ref{fig:reco_vs_true_pi0} shows the result of the fitting, giving $E_{\rm reco} = 0.802 \pm 0.006 \times E_{\rm true}$. This leads to the definition of a corrected energy: $E_{\rm corr} =  E_{\rm reco} / 0.802$.
\par Different methods for applying a bias correction were investigated, partially to address the possibility of an energy-dependent bias correction. While different approaches, which included allowing the fit intercept to float, led to bias corrections which were statistically significant, they ultimately caused an $\mathcal{O}$(1\%) difference in the reconstructed di-photon invariant mass ($M_{\gamma\gamma}$) distribution. While in this work an energy-independent correction is applied, in the future, and depending on the details of the implemented reconstruction, adopting an energy-dependent correction may be beneficial. We note that the bias correction obtained for this reconstruction is smaller than the value of $1/0.70$ obtained from MicroBooNE's previous study of EM activity at lower energy based on Michel electrons~\cite{bib:michel}. This is a consequence of an improved energy reconstruction and refined charge collection capabilities.

\subsection{Energy Resolution Measurement}
\par
\begin{figure}[h] 
  \centering
  \includegraphics[width=0.6\textwidth]{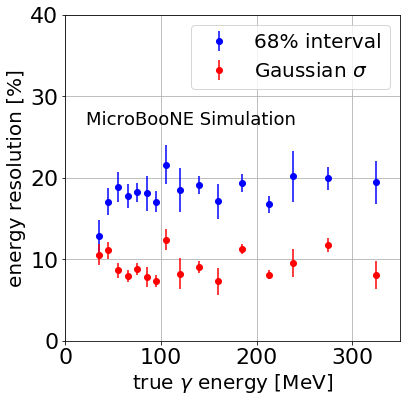}
  \caption{Energy resolution for reconstructed $\gamma$ showers, measured after applying bias corrections. Red points show the fitted $\sigma$ of the Gaussian plus exponential fit function, used to model energy losses and smearing. Blue points denote the 68\% interval half-width of the entire Gaussian plus exponential distribution. Error bars on the measurements in red are obtained from the statistical uncertainty of the fit on the parameter $\sigma$. Error bars on the blue points are obtained from the statistical uncertainty of the fit on the exponential component of the Gaussian plus exponential fit function.}
  \label{fig:eresolution_vsdune}
\end{figure}

 After applying the energy bias correction described above, the fractional energy resolution is again fit to a Gaussian plus exponential-tail function. The energy resolution is quantified in two ways: the Gaussian $\sigma$ of the fit function is taken to represent the resolution for the bulk of the distribution, while the reported 68\% interval accounts for the low-end tail by integrating 68\% of entries asymmetrically around the peak in proportion to the ratio of areas below and above the peak. Measurements of the energy resolution as a function of true photon energy from simulation are reported, using these two definitions, in figure~\ref{fig:eresolution_vsdune}. The fitted Gaussian gives a resolution of 8-12\%, while the 68\% interval method results in a width of 15-20\%. The flat nature of the energy resolution as a function of energy indicates that rather than being limited by the $1/\sqrt{E}$ dependence typical of a total absorption calorimeter, we are in a regime where other effects, including clustering inefficiencies, dominate the energy smearing. This resolution meets, in the \unit[50-300]{MeV} energy range studied, the requirements for electron energy measurements of DUNE (2\% $\oplus$ 15\% / $\sqrt{E \, [{\rm GeV}]}$)~\cite{bib:DUNE} which corresponds to a resolution of 47\% and 27\% at 100 and \unit[300]{MeV}, respectively.

\subsection{$\pi^0$ Energy Resolution}
\par
\begin{figure}[ht] 
\centering
\includegraphics[width=0.6\textwidth]{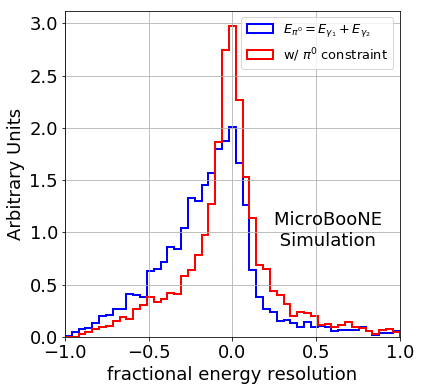}
\caption{Fractional energy resolution, defined as $[E_{\rm reco} - E_{\rm true}] / E_{\rm true}$, for the $\pi^0$ energy computed as the sum of the two photon energies (blue), and employing equations~\ref{eq:epi01} and~\ref{eq:epi02} (red). The $\pi^0$s in this Monte Carlo simulation sample come from simulated $\nu_{\mu}$ CC interactions and have momenta of up to a few hundred MeV.}
\label{fig:epi0}
\end{figure}

 We next study the $\pi^0$ energy resolution. To do so, we use two definitions for energy: in one, the $\pi^0$ energy is given by summing the energy of the two photons, after applying the corrections described in section~\ref{sec:ereco}, while in the second, we make use of the kinematic constraint which can be leveraged assuming the two showers are produced by a $\pi^0$ decay and employing the reconstructed opening angle in the momentum and hence energy determination. The more complex energy definition is shown in equations~\ref{eq:epi01} and~\ref{eq:epi02}, where $\theta$ is the $\gamma\gamma$ opening angle, and $\alpha$ the energy asymmetry between the two showers, defined as $|E_1-E_2|/(E_1+E_2)$.
\begin{eqnarray}
p_{\pi^0} &=& M_{\pi^0} \sqrt{\frac{2}{(1-\alpha^2)(1-\cos\theta)}}, \label{eq:epi01} \\
E_{\pi^0} &=& \sqrt{M_{\pi^0}^2 + p_{\pi^0}^2}.
\label{eq:epi02}
\end{eqnarray}
The first method, which simply integrates the reconstructed energy of the two photons, is susceptible to the lossy biases of shower energy reconstruction presented earlier in this section. While this method leads to a reasonably accurate energy determination, it presents a large negative tail, as shown in the blue distribution of figure~\ref{fig:epi0}. The second method, which makes use of the $\pi^0$ decay kinematic constraint and relies on the energy asymmetry, rather than on the absolute energy, is less sensitive to energy biases, and provides a more accurate and less biased energy resolution, as shown in the red curve. The central peak of this distribution, when fitted to a Gaussian, gives a 10\% resolution on the $\pi^0$ energy. 

\section{Measurements of $\pi^0$ Reconstructed Variables in Data}
\label{sec:pi0}
\par In this section, we present various measurements pertaining to $\pi^0$ and photon reconstruction which are useful to assess the energy calibration of the detector, as well as study electron-photon separation. These results showcase both an accurate modeling of the detector and the capabilities of a robust and sophisticated reconstruction. While the study of $\nu$-Ar interactions is not the subject of this work, we acknowledge that modeling discrepancies in our simulation can contribute to data-MC disagreement. To that end, we limit this work to distributions which are least sensitive to such effects, and present them area-normalized. For graphs in this section blue points are from data, drawn with statistical error bars. Solid lines come from MicroBooNE's BNB simulation and are separated into signal events associated with $\pi^0$ interactions in red and backgrounds with no final-state $\pi^0$ in black. Off-beam backgrounds, subdominant to those associated with neutrino-induced interactions, were determined to not impact the results presented, and are not included in this analysis. Data and simulation comparisons are area-normalized.

\subsection{Reconstructed $\pi^0$ Mass} 
\begin{figure}[ht] 
  \centering
  \includegraphics[width=0.6\textwidth]{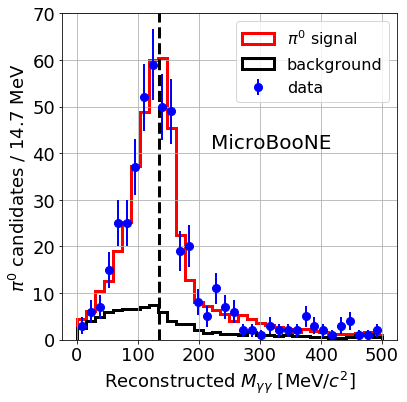}
  \caption{Reconstructed $M_{\gamma\gamma}$ from candidate $\nu_{\mu}$ CC $\pi^0$ events after applying photon-shower energy corrections derived from simulation with an additional 5.5\% shift of the energy in the simulation as explained in the text. The $\chi^2$/bin for the area-normalized data and simulation distributions is 36.1/34. The dashed line denotes the $\pi^0$ mass of \unit[135]{MeV/$c^2$}. The data corresponds to $1.6\times10^{20}$ POT recorded from the BNB.}
  \label{fig:pi0mass}
\end{figure}

Applying the energy reconstruction and bias corrections described in section~\ref{sec:ereco}, we can use the sample of selected di-photon candidates to reconstruct the di-photon mass $M_{\gamma\gamma}$. The mass is obtained from the decay kinematics of the two $\gamma$ showers, through the expression $M_{\gamma\gamma} = \sqrt{2 E_1 E_2 \left( 1 - \cos\theta \right)}$, with $E_1$ and $E_2$ the energy of the two photons, and $\theta$ the reconstructed angle between them. The reconstructed mass is shown in figure~\ref{fig:pi0mass}. Energy corrections are derived from simulation studies on single photon showers, as presented in section~\ref{sec:ereco}. After these corrections, we find good data-simulation agreement in the reconstructed $M_{\gamma\gamma}$ with a $\chi^2$/d.o.f. of $44.7/34$. When we rely on the reconstructed $M_{\gamma\gamma}$ distribution itself as a calibration, as described in Appendix~\ref{app:calib}, we find that a scaling of the simulation by 1.055 (or 5.5\%) relative to data leads to an agreement quantified as 36.1/34 $\chi^2$/d.o.f. with the scaling factor range [0.98, 1.13] encompassing an interval iin $\Delta \chi^2$/d.o.f. of 1.0. The fact that the calibration relying on the invariant mass itself is consistent with the calibration procedure performed relying on stopping muons (found to be accurate at the 3\% level, with possible additional angular dependences which can impact showers more significantly) is an indication of a sound calibration procedure and well modeled detector response for EM showers. In figure~\ref{fig:pi0mass}, and subsequently in comparisons of d$E$/d$x$ distributions, for simulation distributions the 1.055 scaling described in appendix~\ref{app:calib} is applied. Finally, the fact that the reconstructed mass in data lines up with the expected $\pi^0$ mass of \unit[135]{MeV/$c^2$} is an indication that the bias corrections correctly account for the impact on energy reconstruction of the lossy processes of thresholding and under-clustering.

\subsection{Photon d$E$/d$x$}
\label{sec:dedxpi0}
\par The sample of $\gamma$ showers obtained from $\nu_{\mu}$ CC $\pi^0$ interactions is well suited to studying the topological and calorimetric separation of electrons from photons via the measurement of the energy deposition in the initial segment of the shower. We measure shower d$E$/d$x$ using hits deposited in the first \unit[4]{cm} from the photon showering point (in a similar way as presented by the ArgoNeuT collaboration~\cite{bib:argoneutdedx}) and measuring their median d$E$/d$x$ value. This work shows a measurement of shower d$E$/d$x$ performed in a fully automated way for the first time. Due to inefficiencies and biases in calorimetry at small angles with respect to the wire-pitch, we limit ourselves to photon showers that are at an angle with respect to the collection-plane wire direction of at least 33 degrees. In future work, which will incorporate improvements in signal processing already developed by the MicroBooNE collaboration~\cite{bib:SP1,bib:SP2}, shower d$E$/d$x$ information in the full phase-space will be utilized relying, when beneficial, on calorimetric information from the induction planes. Finally, to enhance the purity of the photon sample, only showers with reconstructed energy greater than \unit[50]{MeV} are included in this study.
\par
\begin{figure}[h] 
  \centering
  \includegraphics[width=0.6\textwidth]{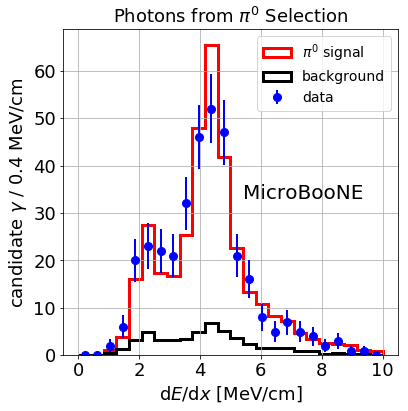}
  \caption{d$E$/d$x$ for photon candidates from selected $\nu_{\mu}$ CC $\pi^0$ events. The $\chi^2$/bin for the area-normalized data and simulation distributions is 15.8/21. The data corresponds to $1.6\times10^{20}$ POT recorded from the BNB.}
  \label{fig:dedx}
\end{figure}

 The reconstructed d$E$/d$x$ is shown in figure~\ref{fig:dedx}. The bulk of the distribution is peaked at \unit[4]{MeV/cm}, expected for a twice minimally ionizing converting photon signature in liquid argon. The peak at \unit[2]{MeV/cm} is a contribution from misreconstruction as well as irreducible backgrounds made up of Compton scattering photons and pair-conversions to an asymmetric $e^{+}e^{-}$, both of which are especially dominant at low $\gamma$ energy. This work shows that MicroBooNE can identify photons by relying on d$E$/d$x$ measured in the first few centimeters of the EM shower development. In this work photons reconstructed with a d$E$/d$x$ below \unit[3.0 (3.5)]{MeV/cm} make up 21\% (30\%) of the selected photons in data. Photon d$E$/d$x$ reconstruction is found to be particularly challenging at low photon energies, due to a combination of the above effects.

\subsection{Photon Conversion Distance}
\par
\begin{figure}[ht] 
  \centering
    \begin{subfigure}[b]{0.48\textwidth}
    \centering
    \includegraphics[width=1.00\textwidth]{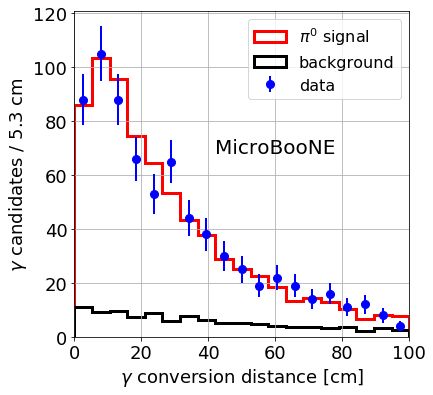}
    \caption{}
    \end{subfigure}
    \begin{subfigure}[b]{0.48\textwidth}
    \centering
    \includegraphics[width=1.00\textwidth]{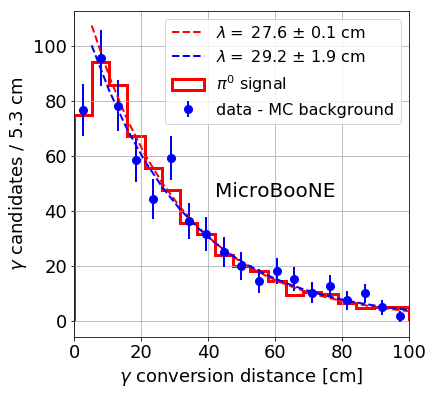}
    \caption{}
    \end{subfigure}
  \caption{(a) Measured conversion distance for reconstructed $\gamma$ showers belonging to the $\nu_{\mu}$ CC $\pi^0$ sample. The $\chi^2$/bin for the area-normalized data and simulation distributions is 15.6/19. The data corresponds to $1.6\times10^{20}$ POT recorded from the BNB. (b) Background subtracted conversion distance distributions fit to an exponential. The exponential constant, denoted as $\lambda$, is found to be \unit[$27.6 \pm 0.1$]{cm} in simulation (red dotted curve) and \unit[$29.3 \pm 1.9$]{cm} in data (in blue), in good agreement with each other. The first bin is not included in the fit. Backgrounds as estimated from simulation as in (a) are subtracted identically in both data and simulation.}
  \label{fig:radlen}
\end{figure}

 An additional variable of interest for photon identification is the photon conversion distance: the separation between the point at which the photon is produced and the point at which it manifests itself in the detector by first contributing to energy deposition. Figure~\ref{fig:radlen} shows the distribution for reconstructed photons from the $\pi^0$ selection. The signal exhibits an exponential behavior as expected for photons converting in the detector. An exponential fit to the background-subtracted distributions, shown in figure~\ref{fig:radlen} (b), results in an extracted conversion distance of \unit[$29.3 \pm 1.9$]{cm} in data. While this measurement does not correct for a conversion distance dependent efficiency, it is consistent with that expected for EM showers of this energy range. 

\section{Conclusions}
\label{sec:conclusions}
\par MicroBooNE has presented a description of a method for the reconstruction of EM interactions in LArTPC detectors in the tens to a few hundred MeV energy range, which are particularly diffuse and stochastic in nature. Particular emphasis has been given to describing the implementation of an energy calibration procedure for EM showers above the Michel electron threshold, identifying and quantifying the primary contributions to energy biases associated with clustering and thresholding. We have shown that we are able to obtain energy resolutions in the range $10 \textrm{--} 20\%$, meeting the requirements for electron energy reconstruction in DUNE's long-baseline neutrino oscillation program~\cite{bib:DUNE}. This reconstruction is used to perform a selection of $\nu_{\mu}$ CC interactions with final state $\pi^0 \rightarrow \gamma\gamma$ decays, leading to 440 candidate events. The reconstructed $M_{\gamma\gamma}$ distribution shows good data-simulation agreement, validating both the calibration and reconstruction being presented. The sample of $\gamma$ EM showers is further used to probe valuable electron-photon separation metrics such as the photon conversion distance and shower d$E$/d$x$, demonstrating the ability to use this information in MicroBooNE for physics analyses.

\appendix

\clearpage

\section{Energy calibration through the $M_{\gamma\gamma}$ $\pi^0$ mass and data-simulation agreement}
\label{app:calib}

\par After applying the calorimetric energy reconstruction and shower energy bias corrections described in section~\ref{sec:ereco}, good agreement is found for the reconstructed $M_{\gamma\gamma}$ invariant mass, both between data and simulation, as well as between the observed and expected reconstructed $\pi^0$ mass value of \unit[135]{MeV/$c^2$}. This section aims to quantify the level of data-simulation agreement, and utilize the reconstructed $M_{\gamma\gamma}$ from $\pi^0$ candidates as an additional, orthogonal, calibration tool.

\begin{figure}[ht]
  \centering
  \includegraphics[width=0.4\textwidth]{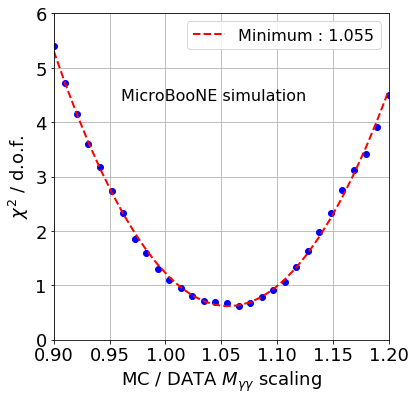}
  \caption{Statistical $\chi^2$ obtained when comparing the area-normalized $M_{\gamma\gamma}$ distribution for data and simulation, as a function of the scaling factor to the data energy scale. An offset of 5.5\% is found, with an interval of $\Delta\chi^2 = 1$ encompassing the scaling range [0.98, 1.13].}
  \label{fig:chi2}
\end{figure}

To quantify the level of agreement in the reconstructed $M_{\gamma\gamma}$ variable, we measure the $\chi^2$ between the area normalized data and simulation distributions, accounting for statistical uncertainties only, as a function of a relative energy scaling applied. Furthermore, the $\chi^2$ is computed only for bins below \unit[250]{MeV/$c^2$} to focus on signal events. The result of this measurement is shown in figure~\ref{fig:chi2} and indicates a 5.5\% discrepancy in energy scale. Specifically, the simulation distribution underestimates by 5.5\% what is reconstructed in data. The $\Delta\chi^2 = 1$ scaling factor correction covers the range [0.98, 1.13]. This factor is compatible with the calibration outlined in section~\ref{sec:ereco} when accounting for the 3\% expected level of uncertainty. 
\par
\begin{figure*}[b]
  \centering
    \begin{subfigure}[b]{0.32\textwidth}
    \centering
    \includegraphics[width=1.00\textwidth]{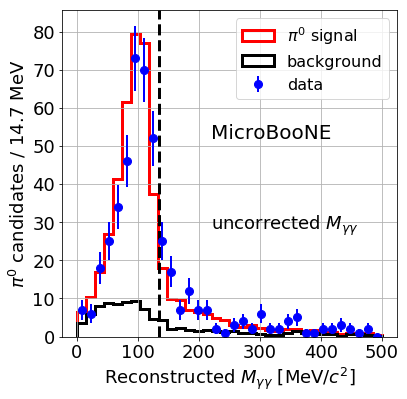}
    \caption{}
    \end{subfigure}
    \begin{subfigure}[b]{0.32\textwidth}
    \centering
    \includegraphics[width=1.00\textwidth]{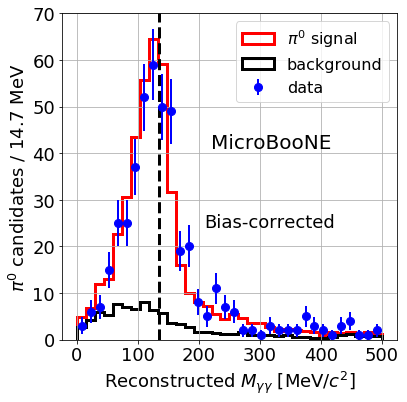}
    \caption{}
    \end{subfigure}
    \begin{subfigure}[b]{0.32\textwidth}
    \centering
    \includegraphics[width=1.00\textwidth]{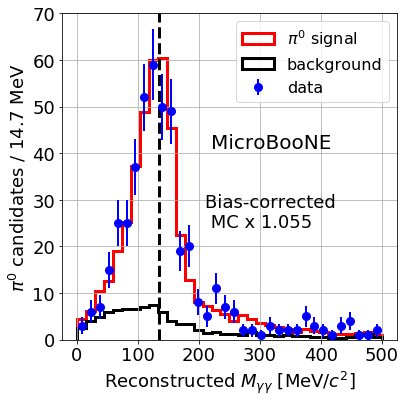}
    \caption{}
    \end{subfigure}
  \caption{Reconstructed $M_{\gamma\gamma}$. (a) After proper calorimetric energy calibration but before any shower energy reconstruction bias corrections. (b) After bias corrections are applied identically in data and simulation. (c) After applying a further 5.5\% correction to simulation. The distribution in (b) has a $\chi^2$/bin of 44.7/34, and the one in (c) gives a value of 36.1/34.}
  \label{fig:pi0mass_rawcorr}
\end{figure*}

 Figure~\ref{fig:pi0mass_rawcorr} (a) and (b) show the reconstructed $\pi^0$ mass distribution obtained without and with bias corrections, applying an otherwise an identical calibration procedure between data and simulation. Figure~\ref{fig:pi0mass_rawcorr} (c) shows the distribution accounting for the remaining 5.5\% discrepancy observed, where simulation has been scaled by a factor of 1.055. The $\chi^2$/d.o.f. moves from $44.7/34$ to $36.1/34$ before and after applying the 1.055 scale factor.

\par
\begin{figure*}[ht] 
  \centering    
    \begin{subfigure}[b]{0.43\textwidth}
    \centering
    \includegraphics[width=1.00\textwidth]{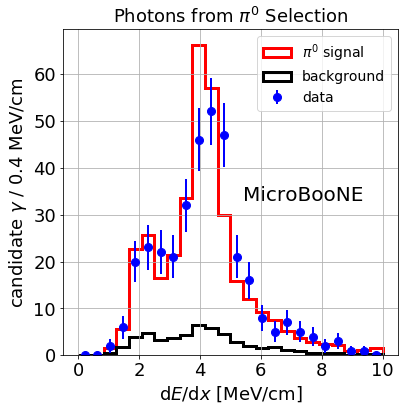}
    \caption{}
    \end{subfigure}
    \begin{subfigure}[b]{0.43\textwidth}
    \centering
    \includegraphics[width=1.00\textwidth]{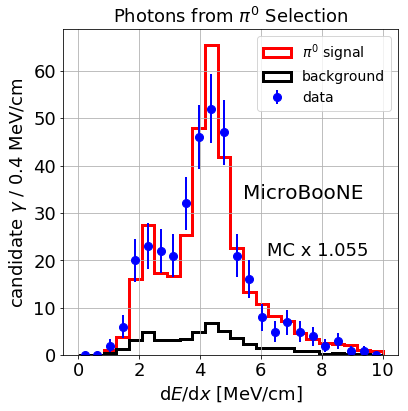}
    \caption{}
    \end{subfigure}
  \caption{d$E$/d$x$ for photon candidates from the $\pi^0$ selection. d$E$/d$x$ is calibrated following the same calorimetry reconstruction described in section~\ref{sec:ereco}, omitting the shower energy-bias corrections which are not relevant here. (a) Reconstructed distribution after applying an identical calibration procedure to data and simulation. (b) Reconstructed distribution after accounting for the additional 5.5\% bias measured from the reconstructed $\pi^0$ mass distribution. The distribution in (a) gives a $\chi^2$/bin of 22.9/21, while for the one in (b) the $\chi^2$ is 15.8/21.}
  \label{fig:dedx_rawcorr}
\end{figure*}

 Finally, the impact of the 1.055 offset on the reconstruction of shower d$E$/d$x$ is shown in figure~\ref{fig:dedx_rawcorr} (a) without and (b) with the 5.5\% correction applied. For this distribution we also see an improved level of data-simulation agreement after the scaling obtained from the $\pi^0$ mass is applied. It is important to note that unlike for shower energy reconstruction, calorimetric d$E$/d$x$ is not impacted by the effect of hit-thresholding and charge-clustering, suggesting that the uncertainty in energy scale for $\gamma$ reconstruction is primarily contributed by the uncertainty in the absolute energy scale calibration. 

\end{document}